\documentclass[prb,twocolumn,showpacs]{revtex4}
\usepackage{graphicx}
\usepackage{multirow}
\usepackage{dcolumn}
\usepackage{bm}
\usepackage[normalem]{ulem}
\usepackage{amsmath}
\usepackage{amsfonts}
\usepackage{amssymb}
\usepackage{color}
\usepackage{colordvi}
\usepackage{dcolumn}
\usepackage{subfigure}

\newcommand{\bq}{\mathbf{q}}

\newcommand{\bA}{\mathbf{A}}

\begin{document}

\title{Vortices and charge order in high-$T_{c}$ superconductors}
\date{\today }
\author{M. Einenkel$^{1,4}$, H. Meier$^{2}$, C. P\'{e}pin$^{3}$, and K. B. Efetov$%
^{1,4}$}
\affiliation{$^{1}$Institut f\"{u}r Theoretische Physik III, Ruhr-Universit\"{a}t Bochum,
44780 Bochum, Germany\\
$^{2}$Department of Physics, Yale University,~New Haven, Connecticut 06520,
USA \\
$^{3}$IPhT, L'Orme des Merisiers, CEA-Saclay, 91191 Gif-sur-Yvette, France\\
$^{4}$National University of Science and Technology \textquotedblleft
MISiS\textquotedblright , Moscow, 119049, Russia}

\begin{abstract}
We theoretically investigate the vortex state of the cuprate high-temperature superconductors in the presence of magnetic fields. Assuming the recently derived nonlinear
$\sigma$-model for fluctuations in the pseudogap phase, we find that the vortex cores consist of two crossed regions of elliptic shape, in which a static charge order emerges.
Charge density wave order manifests itself as satellites to the
ordinary Bragg peaks directed
along the axes of the reciprocal copper lattice. Quadrupole density wave (bond order) satellites, if seen, are predicted to be along the diagonals.  The intensity of the satellites should grow linearly with the magnetic field, in agreement with the result of recent experiments.
\end{abstract}

\maketitle

%
%

\section{Introduction}

Since their discovery\cite{mueller} in 1986, high-temperature~(high-$T_{c}$)
superconductors remain one of the most interesting fields of research in modern
condensed-matter physics. In particular, the origin of the pseudogap (PG) phase, appearing below a temperature $T^\ast$ of the order of a few $100\ \mathrm{K}$, remains one of their most enduring mysteries. The field has
revived recently through a number of spectacular experimental findings,\cite{wise,davis,yazdani,wu,ghiringhelli,chang,achkar,leboeuf,blackburn}
which all give evidence to the presence of charge patterns inside the PG phase.
Understanding the properties of these charge-ordered phases, competing or
coexisting with superconductivity (SC), may significantly help to clarify the physical origin of the PG phase.

While a stripe order combining both charge and spin modulations first
predicted theoretically in Refs.~\onlinecite{zaanen,machida,poilblanc} has been
known for a long time to exist in La-compounds,\cite{tranquada,emery,kivelson,white} the first observation of a modulated structure
in Bi$_2$Sr$_2$CaCu$_2$O$_8$ (BSCCO) was reported only in 2002 by J.~Hoffman \emph{et al.}\cite{hoffman} who
subtracted the scanning tunneling microscope~(STM) response with and
without an applied magnetic field and thus unearthed a checkerboard charge order inside the vortex cores. At the time, it was still  interpreted in terms of the stripes similar to
those La-compounds. It was found that the radius of the region where this
order appears is larger than the radius of the vortex cores.
The close connection between the development of an ordered state and the formation of vortices due to an applied magnetic field has been confirmed by a number of experiments.\cite{Ishida,Levy}
It has recently become clear that the high-$T_c$ compounds of BSCCO and YBa$_2$Cu$_3$O$_{7-x}$ (YBCO)  feature a checkerboard-type charge modulation with wave vectors along the bonds of the CuO lattice.\cite{wu,ghiringhelli,chang,achkar,leboeuf,blackburn,cominxray}

Further, it is well known that under application of a strong  magnetic
field exceeding $17$ T, a striking reconfiguration of the Fermi surface is
observed.\cite{Taillefer1, Taillefer2} After an intense debate, a consensus
emerged in which the reconfiguration of the Fermi surface is attributed to ordering
in the charge sector  with precisely the same
wave vector as the one observed in STM and x-rays.\cite{sebastian1,
sebastian2}(There is no ordering in the spin sector.) Signatures of charge order have also been seen in  magnetic fields above $%
17 $ T in  sound propagation experiments.\cite{leboeuf}

Recent nuclear magnetic resonance (NMR) experiments
 on the cuprate YBCO inside the (hole) doping region $0.11<p<0.12$ for the vortex state showed a charge
modulation  in the core of the vortices.\cite{julien} On the other hand, the
authors of Ref.~\onlinecite{chang} studied YBCO in a magnetic field at hole doping $p=0.12$ using
the high energy x-ray scattering technique. They found the charge order
not only in the pseudogap phase below an ordering temperature $T_{\mathrm{cdw%
}}<T^{\ast }$, but also in the superconducting state with the maximum
magnitude of the charge density wave (CDW) at the superconducting transition
temperature $T_{c}$. Remarkably, below $T_{c}$ the lattice modulation peak
intensity grows linearly as a function of the magnetic field.

We can conclude from all these experiments that a charge-order state
competes with the superconducting state in high-$T_{c}$ cuprates and appears or is enhanced by a moderate magnetic field destroying or
suppressing the superconductivity.

Recently, several of us have suggested to describe this competition between $d$-wave superconductivity and charge order in terms
of a two dimensional $\mathrm{O}(4)\times \mathrm{O}(4)$ symmetric nonlinear $%
\sigma $-model.\cite{emp} The components of the unit vectors represent fluctuating order parameters for superconductivity and a
charge-modulated state. In this theory, the PG
state is a fluctuating composite state made of superconducting  and
charge  suborders corresponding to the disordered phase of the $%
\sigma $-model. The magnetic field can naturally be taken into account
within this $\sigma $-model, and the competition between superconductivity and charge order can be explained and described using a renormalization-group scheme.\cite{mepe} The results obtained in such a study are in good agreement with the
results of the experiment on sound propagation.\cite{leboeuf}

The $\sigma $-model approach of Refs.~\onlinecite{emp} and \onlinecite{mepe} was further
developed by L. Hayward \emph{et al.}\cite{ssscience,sachdevmag} who studied an $\mathrm{O}(6)$ symmetric $\sigma$-model that is identical to the $\mathrm{O}(4)\times \mathrm{O}(4)$  model if the superconducting phases of the two $\mathrm{O}(4)$  sectors are
interlocked. Comparing the results of a Monte-Carlo simulation based on this $\sigma$-model with experimental data on x-ray scattering, they reproduced the observed temperature dependency of the charge-order signal.

The nonlinear $\sigma $-model of Ref.~\onlinecite{emp} has been derived
starting from the so-called spin-fermion model.\cite{ac,acs} The charge order produced by this approach has been identified as a quadrupole density wave~(QDW)
with  modulation vectors directed along the diagonals of the CuO lattice.
The instability toward the charge modulation with this symmetry had been
discussed earlier\cite{ms2} under the name \textquotedblleft valence bond
solid\textquotedblright . However, the fact that experiments so far see a primary charge order along the CuO bonds has been troubling theoreticians for years.

The CDW order with the modulation vectors directed along the bonds has been
addressed by us in another publication\cite{mepe2} followed by several
other proposals.\cite{wc,allais,chowdhury,allais2} In Ref.~\onlinecite{mepe2}, the
CDW order is considered as a corollary attribute of the QDW/SC order,
co-existing with it in the PG phase for $T<T_{\mathrm{cdw}}$ or in a strong
magnetic field inside the superconducting phase. The CDW is only a
by-product of the SC/QDW order, induced by superconducting fluctuations,  so that the shape of the transition lines in
the $T$-$B$ phase diagram, or the structure of the vortices, is determined by
the competition between the SC and QDW suborders inside the PG phase.

Due to its unusual structure, it is not easy to observe the
QDW order directly using non-resonant x-ray scattering. At the same time, the
x-ray scattering may get a weaker signal from the secondary CDW order,
and one can speak of studying the SC/QDW competition using the x-ray
technique. For a $\sigma$-model description of the competition between SC and the experimentally observed CDW, x-ray scattering would probe this competition
directly.
\\

In this work we use the nonlinear $\sigma$-model\cite{emp} to investigate
the structure of a quantum vortex in the superconducting phase of a
high-$T_{c}$ superconductor. Previous works concentrated on general
properties of the phase diagram without magnetic fields\cite{emp,mepe2} or
the transition from a uniform superconducting state into a uniform charge-ordered state under the influence of a strong magnetic field.\cite{mepe} The properties of the vortex phase itself have remained an open question and are
addressed by this work. We do not try to clarify the nature of the charge order  here,
assuming that it can be probed by different methods including x-ray
scattering and STM spectroscopy.

For this purpose, we use the generalization of the nonlinear $\sigma$-model
with a magnetic field introduced in Ref.~\onlinecite{mepe}.  Based on this model, we derive equations for
the order parameter describing a quantum vortex carrying one magnetic flux
quantum. We show that the symmetry of this order parameter leads to charge
ordering inside the vortex core.

In a second step, we argue that this order is visible in x-ray scattering
experiments by contributing to satellite peaks close to the standard Bragg
peaks. Finally, the position of this satellites allows us to distinguish
between the different types of charge ordered states, QDW or CDW. We show
that the modulation peak intensity should be proportional to the magnetic
field, which is in agreement with the results of the experiment.\cite{chang}

\section{Composite order parameter and main equations}

\label{secII}

\subsection{Nonlinear $\protect\sigma$-model}

\begin{figure}[t]
\centerline{\includegraphics[width=\linewidth]{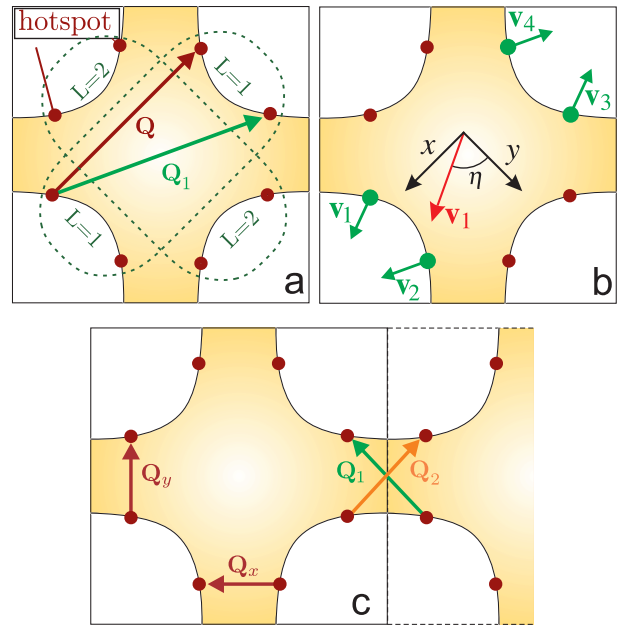}}
\caption{(Color online) (a) Brillouin zone and hotspots connected with the antiferromagnetic wave vector $\mathbf{Q}$. Hotspots can be organized in terms of two quartets $L=1$ and $L=2$. (b) The hotspots forming the $L=1$ quartet are indicated, and the velocities are defined via the angle $\eta$. (c) Wavevectors of the different orderings are presented. QDW order is modulated with $\mathbf{Q}_{1,2}$ while the CDW is modulated with $\mathbf{Q}_{x,y}$.}
\label{Fig1}
\end{figure}

Below a temperature~$T^{\ast }$, the spin-fermion model features a PG
phase \cite{emp} characterized by an order parameter comprising both $d$%
-wave superconductivity and a charge order. Although the direct derivation leads to
superconductivity competing with the QDW, the final
equations can phenomenologically also be used for  other types of charge order. As
discussed in the Introduction, an additional charge order may be bound to the
QDW, which allows one to observe the latter indirectly by, e.g., an x-ray
technique. Thus, we consider explicitly the model with the SC/QDW
composite order parameter as in Refs.~\onlinecite{emp,mepe} but
having in mind also some broader applications, when discussing the symmetry
of the charge order.

The order parameter may be represented in the form of an $\mathrm{SU}(2)$
unitary matrix in the particle-hole (Gorkov-Nambu) space,
\begin{equation}
u^{L}=\left(
\begin{array}{cc}
\Delta _{\mathrm{QDW}}^{L} & \Delta _{\mathrm{SC}}^{L} \\
-\Delta _{\mathrm{SC}}^{L\ast } & \Delta _{\mathrm{QDW}}^{L\ast }%
\end{array}%
\right) \ .  \label{matU}
\end{equation}%
By unitarity, the parameters for the superconducting and charge orders, $%
\Delta _{\mathrm{SC}}^{L}$ and $\Delta _{\mathrm{QDW}}^{L}$ are subject to
the nonlinear constraint $|\Delta _{\mathrm{QDW}}^{L}|^{2}+|\Delta _{%
\mathrm{SC}}^{L}|^{2}=1$. It is ultimately this constraint that leads to
rather unusual superconducting properties of the system. The upper index~$%
L=1,2$ in Eq.~(\ref{matU}) refers to the two quartets of hotspots,
connected within the Brillouin zone by either $\mathbf{Q}=(\pi ,\pi )$ for $%
L=1$ or $(-\pi ,\pi )$ for $L=2$, see Fig.~\ref{Fig1}(a). In the
hotspot-only approximation,\cite{ms2,emp} these two quartets and their
order parameters are decoupled.

Superconductivity and charge order are degenerate suborders of the pseudogap
order described by Eq.~(\ref{matU}). Finite curvature of the Fermi surface
or a magnetic field lift the degeneracy below a temperature~$T_c$, the
former to favor superconductivity and the latter to support the charge
order. At higher temperatures, $T_c<T<T^*$, thermal fluctuations eventually
restore the degeneracy. As these thermal fluctuations play a central role in
our study, let us briefly discuss the nonlinear $\sigma$-model describing
them. For an extensive discussion and derivation, we refer the reader to
Ref.~\onlinecite{emp}.

We are interested in the limit of long-wavelength thermal fluctuations,
described by $u_{\mathbf{q}}^{L}=u_{0}^{L}+\delta u_{\mathbf{q}}^{L}$ on
small momenta~$\mathbf{q}$. The fluctuation modes arise from coupling to the
electrons; see Fig.~\ref{Fig1}(b). Let us consider the order parameter~$%
u^{L=1}$, Eq.~(\ref{matU}), for the first quartet. The order parameter~$%
u^{L=1}$ coherently couples both hotspot~$1$ with~$3$ and hotspot~$2$ with~%
$4$. These pairs of hotspots are effectively nested, i.e. $\mathbf{v}_{1}=-%
\mathbf{v}_{3}$ and $\mathbf{v}_{2}=-\mathbf{v}_{4}$. As a result, the
contribution to fluctuation modes due to the first pair of hotspots allows
a momentum dependence only in the form of $(\mathbf{v}_{1}\mathbf{q})^{2}$,
whereas the second pair leads to a dependence only on~$(\mathbf{v}_{2}%
\mathbf{q})^{2}$. Being gapless, the fluctuation modes are thus described in
the leading order by the free-energy functional~$F_{L=1}=T\mathcal{F}_{L=1}$
with
\begin{align}
& \mathcal{F}_{1}[\delta u^{1}]  \label{h01} \\
& \quad =\frac{1}{2tv^{2}}\int \frac{d^{2}\mathbf{q}}{(2\pi )^{2}}\big\{(%
\mathbf{v}_{1}\mathbf{q})^{2}+(\mathbf{v}_{2}\mathbf{q})^{2}\big\}\mathrm{tr}%
\big[(\delta u_{q}^{1})^{\dagger }\delta u_{q}^{1}\big]\ .  \notag
\end{align}%
The dimensionless coupling constant~$t$ is given by the temperature in units
of the pseudogap scale~$T^{\ast }$, $t=\alpha T/T^{\ast }$, with the
numerical coefficient $\alpha \approx 0.74$.\cite{emp} Choosing the
coordinate axes along the diagonals of the Brillouin zone, see Fig.~\ref{Fig1}(b), we have $\mathbf{v}_{1}=v(\sin \eta ,\cos \eta )$ and $\mathbf{v}%
_{2}=v(\sin \eta ,-\cos \eta )$ with~$v$ the value of the Fermi velocity.
Thus, $(\mathbf{v}_{1}\mathbf{q})^{2}+(\mathbf{v}_{2}\mathbf{q}%
)^{2}=2q_{x}^{2}\sin ^{2}\eta +2q_{y}^{2}\cos ^{2}\eta $. Transforming to
real space, we write the functional~$\mathcal{F}=\mathcal{F}_{1}+\mathcal{F}%
_{2}$ for the free energy of both quartets in the form of two $\sigma $-models
on the manifold~$\mathrm{SU}(2)$,
\begin{equation}
\mathcal{F}[u]=\frac{1}{t}\sum_{L=1}^{2}\int d^{2}\mathbf{r}\ \mathrm{tr}%
\big[{\nabla }^{L}(u^{L})^{\dagger }\nabla ^{L}u^{L}\big]\ .  \label{h02}
\end{equation}%
For~$\eta \neq 45^{\circ }$, the gradients~$\nabla ^{L}$ are anisotropic.
For~$L=1$, we find according to our above analysis
\begin{equation}
{\nabla }^{L=1}=\big(\sin \eta \ \partial _{x},\cos \eta \ \partial _{y}\big)%
\equiv \Gamma _{1}\nabla \ ,  \label{h03}
\end{equation}%
where
\begin{equation}
\Gamma _{1}=\left(
\begin{array}{cc}
\sin \eta & 0 \\
0 & \cos \eta%
\end{array}%
\right) \ .  \label{h03a}
\end{equation}%
For the quartet~$L=2$, obtained by turning the $L=1$ quartet by $90^{\circ }$%
, the gradient reads
\begin{equation}
{\nabla }^{L=2}=\big(\cos \eta \ \partial _{x},\sin \eta \ \partial _{y}\big)%
\equiv \Gamma _{2}\nabla,  \label{h04}
\end{equation}%
with
\begin{equation}
\Gamma _{2}=\left(
\begin{array}{cc}
\cos \eta & 0 \\
0 & \sin \eta%
\end{array}%
\right) \ .  \label{h04a}
\end{equation}%
The different anisotropies in the two $L$ sectors lead to unusual effects in
the geometry of vortices in the presence of a magnetic field, as we will
 show below.

The $\sigma $-model~(\ref{h02}) for the gapless fluctuations of the
pseudogap order parameter has been derived for linear Fermi surfaces around
the hotspots. Taking into account the finite curvature of the Fermi
surface, we have to supplement the model~(\ref{h02}) by the term\cite{emp}
\begin{equation}
\mathcal{F}_{\mathrm{curv}}[u]=\frac{\mu ^{2}}{t}\sum_{L=1}^{2}\int d^{2}%
\mathbf{r}\ \mathrm{tr}\big[\tau _{3}(u^{L})^{\dagger }\tau _{3}u^{L}\big]\ .
\label{h05}
\end{equation}
The coupling constant~$\mu $ has dimension of inverse length and grows with
increasing the curvature. With $\tau _{3}$ denoting the Pauli matrix in
particle-hole space, we see that $\mathcal{F}_{\mathrm{curv}}$ breaks the
symmetry between superconductivity and charge order favoring the
superconducting suborder.

\subsection{Vortex solution}

Let us parametrize the unitary matrix~$u^{L}$ for the order parameter in
Eq.~(\ref{matU}) using polar coordinates. Introducing an \textquotedblleft
angle\textquotedblright ~$\theta ^{L}$ between superconductivity and charge
order, we write
\begin{equation}
\Delta _{\mathrm{QDW}}^{L}=\sin \theta ^{L}\mathrm{e}^{\mathrm{i}\chi
_{L}}\quad \mathrm{and}\quad \Delta _{\mathrm{SC}}^{L}=\cos \theta ^{L}\,%
\mathrm{e}^{\mathrm{i}\phi _{L}}\ .  \label{parametrization}
\end{equation}%
The parameters~$\chi _{L}$ and~$\phi _{L}$ are the phases of the charge and
superconducting order, respectively. Fluctuations of the phase~$\chi _{L}$
of the charge order are relevant close to~$T_{c}$ but negligible in the
regime~$T\ll T_{c}$ we are interested in. We therefore assume that $\chi
_{L}=0$ for both $L=1$ and $L=2$.

We include the magnetic field into the free-energy functional~(\ref{h02}) by
minimal coupling,\cite{mepe}
\begin{equation*}
\nabla ^{L}u^{L}\rightarrow {\nabla }^{L}u^{L}+\frac{\mathrm{i}e}{c}{\mathbf{%
A}}^{L}[\tau _{3},u^{L}],
\end{equation*}%
where ${\mathbf{A}}^{L=1}=(\sin \eta \ A_{x},\cos \eta \ A_{y})$ and ${%
\mathbf{A}}^{L=2}=(\cos \eta \ A_{x},\sin \eta \ A_{y})$ are the reduced
vector potentials due to the anisotropies in the two $L$-sectors.
Furthermore, we add the contribution of the magnetic field to the free
energy in units of temperature,
\begin{equation}
\mathcal{F}_{B}[{\mathbf{A}}]=\frac{1}{T}\int d^{3}\mathbf{r}\ \frac{[\nabla
\times {\mathbf{A}}]^{2}}{8\pi }\   \label{freeB}
\end{equation}%
where $\mathbf{A}=(A_{x},A_{y},0)$.

The $\sigma $-model has been derived for a single plane of CuO, and the
total free energy is obtained by summing the contributions of all individual
layers leading to an anisotropic three-dimensional~(3D) model. Basically,
two kinds of models are commonly used for the description of layered
superconductors. For highly anisotropic systems, \emph{discrete} two-dimensional layers are coupled by Josephson terms giving rise to interesting
behavior of the vortex solution.\cite{blatter} If the anisotropy is not very large, a continuous anisotropic 3D model is applicable.
Actually, for fields perpendicular to the layers, the difference between these two model is not
very important, and under the assumption that the magnetic field varies on length scales much larger than the layer thickness~$d$, integration in the vertical $z$-direction simply yields a factor of $d$ for each individual layer.

In the parametrization~(\ref{parametrization}) of the order parameter, the $%
\sigma$-model~(\ref{h02}) in the presence of the magnetic field reads
\begin{align}
\mathcal{F}[\theta,\phi,\mathbf{A}] &= \frac{2}{t} \sum_{L=1}^2\int d^2
\mathbf{r}\Big\{ \big(\nabla^L \theta^L \big)^2  \notag \\
+ \Big(\nabla^L \phi_L &+ \frac{2 e}{c} {\mathbf{A}}^L \Big)^2\cos^2\theta^L
- \mu^2 \cos 2\theta^L \Big\} \ .  \label{freeend}
\end{align}
For each sector~$L$, this model is reminiscent of the anisotropic
Ginzburg-Landau functional.\cite{blatter} However, the nonlinearity of the
model~(\ref{h02}) becomes noticeable in the $\cos$-dependencies of the field~%
$\theta^L$. Also note that there are two order parameters~$\theta^1$ and~$%
\theta^2$, which in the presence of the magnetic field are coupled. The free
energy functional~(\ref{freeend}) is minimal for $\theta^L$ and $\mathbf{A}$
satisfying the Ginzburg-Landau-type equations
\begin{widetext}
\begin{align}
 \nabla^L \nabla^L  \theta^L + \frac 12 \left(\nabla^L \phi_L + \frac{2e}{c} \bA^L   \right)^2 \sin 2 \theta^L - \mu^2 \sin 2 \theta^L = 0,
         \label{eq1} \\
\left(  \nabla \times \nabla\times  \bA\right)  + \frac{32 \pi eT^*}{c \alpha d } \sum_{L=1}^2 \Gamma_L \left( \nabla^L \phi_L + \frac{2e}{c}  \bA^L   \right)\cos^2 \theta^L = 0,
         \label{eq2}
\end{align}
\end{widetext}
with the ``anisotropy matrices'' $\Gamma_L$ defined in Eqs.~(\ref{h03a})
and~(\ref{h04a}).

In the following, we are investigating the vortex solution for $\theta _{L}(%
\mathbf{r})$, associated with a magnetic flux equal to the flux quantum $%
\Phi _{0}=\pi c/e$.\cite{tinkham} For convenience, we choose the center of
the vortex to be the origin. The anisotropy then suggests for the phases the
spatial dependencies
\begin{equation*}
\phi _{L=1}=\arctan \left( \tan \eta \ \frac{y}{x}\right)
\end{equation*}%
and
\begin{equation*}
\phi _{L=2}=\arctan \left( \cot \eta \ \frac{y}{x}\right) \ .
\end{equation*}%
The different anisotropic behavior for each $L$ sector complicates the
search for an exact solution of Eqs.~(\ref{eq1}) and~(\ref{eq2}). However, a
simplification is possible because the superconducting order parameter and
the magnetic field vary on different length scales. Indeed, $\theta _{L}$
becomes constant for $r=|\mathbf{r}|>\xi _{\mathrm{cor}}$ (with $\xi _{%
\mathrm{cor}}$ denoting the correlation length) whereas the magnetic field
varies on the scale of the penetration depth~$\lambda $, which for a strongly
type-II superconductor is much larger. We will verify the validity of this
assumption \emph{a posteriori} by giving an estimate of the Ginzburg-Landau
parameter $\kappa \equiv \lambda /\xi _{\mathrm{cor}}$ based on our
analysis. Thus, as the first step, we determine the magnetic field using the
fact that far from the vortex core, $\cos \theta ^{L}\simeq 1$; cf. Eq.~(\ref%
{parametrization}). Taking the curl on both sides of Eq.~(\ref{eq2}), we
find that on scales $\gg \xi $,
\begin{equation}
\triangle \mathbf{B} - \lambda ^{-2}\mathbf{B} = -\lambda ^{-2}\Phi _{0}\delta (%
\mathbf{r})\mathbf{e}_{z}\ .  \label{equasB}
\end{equation}%
In terms of the microscopic parameters, the penetration depth~$\lambda $ is
given by $\lambda ^{-2}=64\pi e^{2}T^{\ast }/c^{2}d\alpha $. Symmetry, in
particular the absence of anisotropy, suggests using polar coordinates~$%
(r,\varphi )$. For the vector potential, we choose a gauge such that $%
\mathbf{A}=A(r)\mathbf{e}_{\varphi }$ and $d(rA)/dr=rB$ for $\mathbf{B}%
(r)=B(r)\mathbf{e}_{z}$. Proper boundary conditions are $A(0)=0$ and $%
A(r)=\Phi _{0}/(2\pi r)$ for $r\rightarrow \infty $, guaranteeing a total
magnetic flux of one flux quantum~$\Phi _{0}$.

With the magnetic field~$\mathbf{B}(\mathbf{r})$ at hand, we are then in the
position to determine the angular fields~$\mathbf{\theta }^{L}(\mathbf{r})$
using Eq.~(\ref{eq1}). For both quartets of hotspots, $L=1$ and $L=2$, we
impose the same boundary conditions $\theta ^{L}(0)=\pi /2$ and $\theta
^{L}(\infty )=0$, in line with the inherent $d$-wave symmetry of the order.
It is convenient to perform for each sector~$L$ the coordinate
transformation $\mathbf{r}=\Gamma _{L}\tilde{\mathbf{r}}$ with $\Gamma _{L}$
defined in Eqs.~(\ref{h03a}) and~(\ref{h04a}). As a result, we have thus
mapped the system of two anisotropic systems to two isotropic ones in $L$%
-dependent coordinates~$\tilde{\mathbf{r}}$.

In spherical coordinates $(\bar{r},\varphi )$, where the dimensionless
coordinate $\bar{r}=|\tilde{\mathbf{r}}|/\xi $ is the radius in units of the
characteristic length $\xi =1/(\sqrt{2}\mu )$, Eq.~(\ref{eq1}) is for each~$%
L $ reduced to
\begin{equation}
\Delta \theta _{L}+\frac{1}{2}\left( \frac{1}{\bar{r}}+\frac{e\sin (2\eta
)\xi ^{2}}{2c}B(0)\bar{r}\right) ^{2}\sin 2\theta _{L}-\frac{1}{2}\sin
2\theta _{L}=0\ .  \label{eq1a}
\end{equation}%
Deriving Eq.~(\ref{eq1a}), we approximated the vector potential as $%
A(r)\simeq B(0)r/2$, which within the scale of the vortex constitutes the
leading order. The second term in the large parentheses is parametrically much smaller
than the first one (because $(e/c)\xi ^{2}B(0)=\kappa ^{-2}\ln \kappa \ll 1$%
\cite{tinkham}), and it may therefore be omitted in the numerical solution.

Vortex solutions to Eq.~(\ref{eq1a}) are rotationally symmetric in the $%
\tilde{\mathbf{r}}$-coordinate system. Transforming back to the physical
coordinates~$\mathbf{r}=(x,y)$, we find for the hotspot quartet~$L=1$ that $%
\theta^1$ depends on coordinates only as a function of~$\sqrt{%
(x/\sin\eta)^2+(y/\cos\eta)^2}$, while for the second quartet, $\theta^{L=2}$
is an effective function of $\sqrt{(x/\cos\eta)^2+(y/\sin\eta)^2}$. Vortices
in the superconducting order parameter have thus the shape of ellipses, with
the ellipses in sectors~$L=1$ and~$L=2$ rotated by $90^\circ$ with
respect to each other.

\section{Vortex State}

Generally, solving Eq.~(\ref{eq1a}) requires numerical methods. Figure~\ref%
{vortexsolution}(a) shows amongst others a plot of $|\Delta _{\mathrm{SC}%
}|=\cos \theta ^{L}$ as a function of the dimensionless radius~$\bar{r}$. In
certain limits, however, we may obtain approximate analytical solutions and
use these solutions to estimate characteristic parameters of the system, such
as the Ginzburg-Landau parameter~$\kappa $. In the superconducting state, $%
\theta ^{L}=0$, while for $\theta ^{L}=\pi /2$ the system shows pure charge
order.

We assume that the system temperature is below~$T_{c}$ so that the system is
a superconductor, which for a sufficiently strong magnetic field is
penetrated by vortices. In the middle of the core of a single vortex, the
superconducting order vanishes. Setting $\theta ^{L}(\mathbf{r})=\pi
/2+\delta \theta ^{L}(\mathbf{r})$, we may then expand the left-hand side of
Eq.~(\ref{eq1a}) in $\delta \theta ^{L}$. As a result, the problem is
reduced to the single vortex in the conventional Ginzburg-Landau theory.\cite%
{tinkham} In particular, we find that the characteristic length parameter~$%
\xi $ determines the size of the vortex core and corresponds to the
correlation length~$\xi _{\mathrm{cor}}$. Thus, expressing the
Ginzburg-Landau parameter~$\kappa $ in terms of the parameters of the model~(%
\ref{freeend}), we find the intermediate result
\begin{equation}
\kappa \sim \kappa ^{\ast }\simeq \sqrt{\frac{\alpha c^{2}d\mu ^{2}}{%
32e^{2}\pi T^{\ast }}}  \label{h06}
\end{equation}%
The parameter~$\mu ^{2}$ characterizing the curvature has been extracted\cite%
{mepe} from the data for the zero-temperature critical magnetic field~$%
B_{c2} $ measured in a sound experiment on YBCO.\cite{leboeuf} According to
a fit in Ref.~\onlinecite{mepe}, $\mu ^{-1}\approx 9\ \mathrm{nm}$. For the
pseudogap temperature, we use $T^{\ast }\approx 250\ \mathrm{K}$ and for the
width~$d$ of a CuO layer the estimate $d\sim 10\ \mathrm{\mathring{A}}$%
. Then, Eq.~(\ref{h06}) yields the rough estimate of $\kappa ^{\ast }\sim 10$
and already confirms the assumption of a rather strongly type-II
superconductor used when making approximations.

\begin{figure}[t]
\centerline{\includegraphics[width=\linewidth]{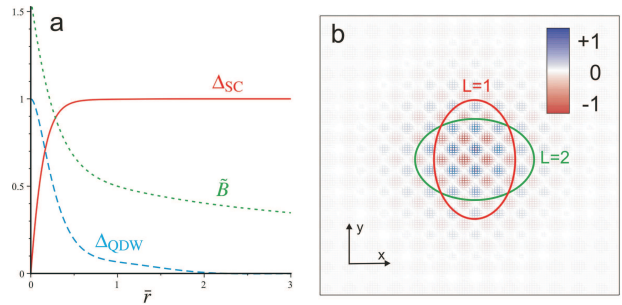}}
\caption{(Color online) (a) Result of numerical studies of Eqs.~(\protect\ref%
{equasB}) and~(\protect\ref{eq1a}) for a single vortex: superconducting
order parameter $\Delta_{\mathrm{SC}}$ (red, solid), charge order $\Delta_{\mathrm{%
QDW}}$ (blue, dashed), and magnetic field $\bar B$ (green, dotted) in arbitrary units as a
function of the dimensionless radius~$\bar{r}$. The nonlinear constraint on
the total order parameter~(\protect\ref{matU}) leads to the rise of charge
order triggered by simultaneous decay of the superconducting order. The
magnetic field penetrates deep into the plane. (b) Checkerboard density wave
order inside the vortex core of radius $\sim\protect\xi$.}
\label{vortexsolution}
\end{figure}

However, a more accurate estimate shows that the correlation length~$\xi _{%
\mathrm{cor}}$ is smaller than the characteristic length~$\xi $ in Eq.~(\ref%
{eq1a}) so that~$\kappa $ becomes even larger. Indeed, based on the
numerical solution for the nonlinear equation for~$\Delta _{\mathrm{SC}}$,
cf. Fig.~\ref{vortexsolution}(a), we extract an estimate for the correlation
length as the length at which $\Delta _{\mathrm{SC}}=1/2$. This procedure
yields $\xi _{\mathrm{cor}}\approx 0.1\xi $. As a result of this refined
analysis, we thus obtain a Ginzburg-Landau parameter
\begin{equation*}
\kappa \approx 10\kappa ^{\ast }\sim 100.
\end{equation*}%
This still rough order-of-magnitude estimate is in line with the literature,%
\cite{blatter} although perhaps somewhat larger than expected. In
particular, it \emph{a posteriori} justifies the approximations done in Sec.~%
\ref{secII}.\bigskip

Let us now dwell on more non-trivial effects resulting from the nonlinear $%
\sigma $-model for matrices~$u^{L}$, Eq.~(\ref{matU}). We have already seen
that the nonlinear constraint $|\Delta _{\mathrm{QDW}}^{L}|^{2}+|\Delta _{%
\mathrm{SC}}^{L}|^{2}=1$ between the superconducting and charge suborders
effectively enhances the inverse correlation length~$\xi _{\mathrm{cor}%
}^{-1} $. More strikingly, this constraint makes charge order
emerge automatically as soon as superconductivity is (locally) suppressed by the magnetic
field.\cite{mepe} As a result, the vortex cores carry charge order.
Moreover, as shown in Fig.~\ref{vortexsolution}, there is a region around a
vortex between the radius scales of $\xi _{\mathrm{cor}}$ and~$\xi $ where
superconductivity co-exists with still appreciable charge order.

In real space, as discussed in Sec.~\ref{secII}, the vortices in each of the
two sectors~$L$ for the two quartets of hotspots have in general elliptic
shapes. Only for the special angle of $\eta =45^{\circ }$, cf. Fig.~\ref%
{Fig1}, do these ellipses turn into circles that for both sectors are the same.
In the general situation, $\eta \neq 45^{\circ }$ and the two sectors~$L$
feature anisotropic vortices that are rotated by $90^{\circ }$ with
respect to each other. As a result, while breaking rotational symmetry, the
geometry of a single vortex reflects the $d$-wave spatial symmetry. Fig.~\ref{vortexsolution}(b) visualizes the checkerboard charge order inside the
two crossed vortices. At the boundaries, where due to the anisotropy only
one of the $L$-sectors still shows vortex features, the density shows a
(very) local stripe structure.

\section{Discussion of x-ray experiments}

Let us now address the question of how the predicted state can be observed
in x\nobreakdash-ray experiments. By conventional hard x\nobreakdash-ray
scattering techniques, one basically measures the Fourier transform of the
charge density. A charge order is thus in principle detectable using such
experiments. Below~$T_{c}$, our theoretical analysis based on the $\sigma $%
-model for the pseudogap state shows the emergence of a charge order inside
vortex cores, provided a sufficiently strong magnetic field creates
vortices. This competing charge order has been identified in Ref.~\onlinecite{emp} as a quadrupole density wave, or equivalently bond
order.\cite{ms2} This order is characterized by two charge density wave
orders on the Cu bonds, where the oxygen atoms are situated. A phase
difference of $\pi $ between O atoms on bonds in $x$-direction and those in $%
y$-direction established the quadrupolar charge order around each Cu atom.

Explicitly, the QDW leads to a charge modulation of oxygen atoms\cite{emp} of
the form
\begin{align}
\rho_{\mathrm{QDW},x/y}(\mathbf{r}) = \pm |\Delta_{\mathrm{QDW}}| \big[ \sin(%
\mathbf{Q}_1\mathbf{r}) + \sin(\mathbf{Q}_2\mathbf{r}) \big] \ .
\label{rhoQDW}
\end{align}
The overall sign is different for  bonds in $x$- and $y$-directions. The
wave vectors~$\mathbf{Q}_1$ and~$\mathbf{Q}_2$ connect opposing hotspots,
see Fig.~\ref{Fig1}(c).

Since at each site in the Cu lattice the average charge density associated
with the QDW modulation of formula~(\ref{rhoQDW}) is zero, the QDW itself seems
 difficult to impossible to be detected in x\nobreakdash-ray
experiments. Also, the wave vector~$\mathbf{Q}_{1}$ and~$\mathbf{Q}_{2}$ are
not the ones observed in either STM\cite{julien,cominSTM} or x\nobreakdash%
-ray\cite{cominxray} experiments, which instead indicate wave vectors~$%
\mathbf{Q}_{x}$ and $\mathbf{Q_{y}}$ along the bonds, cf. Fig.~\ref{Fig1}(c).
At the same time, theoretical approaches that take a microscopic point of
view, assuming for example a model of a single antiferromagnetic quantum
critical point,\cite{ms2,emp,laplaca} typically identify the QDW with wave
vectors $\mathbf{Q}_{1}$ and~$\mathbf{Q}_{2}$ to be the order associated
with the leading instability in the particle-hole channel. Very recently,
theoretical ideas and mechanisms \cite{mepe2,allais,chowdhury} have been
developed that may supplement the QDW picture \cite{ms2,mepe2} by a true
charge density wave order on the Cu atoms with the experimentally
observed wave vectors. These ideas include extensions of the quantum
critical hotspot model,\cite{mepe2} taking into account strong on-site Coulomb interactions,\cite{allais} and non-trivial interplay between charge order and superconducting fluctuations.\cite{mepe2,chowdhury,allais,allais2} It is
not yet clear, though, whether CDW will eventually have to be regarded as
co-existing or competing with the QDW. Taking a phenomenological approach,\cite%
{ssscience, sachdevmag} we choose in the following to discuss the simplest
picture of a nonlinear $\sigma $-model with CDW being the competitor of
superconductivity instead of the QDW. In this case, we assume the following form
of the charge density on the Cu atoms
\begin{equation}
\rho _{\mathrm{CDW}}(\mathbf{r})=|\Delta _{\mathrm{CDW}}|\big[\sin (\mathbf{Q%
}_{x}\mathbf{r})+\sin (\mathbf{Q}_{y}\mathbf{r})\big]  \label{rhoCDW}
\end{equation}%
with the observed wave vectors~$\mathbf{Q}_{x}$ and $\mathbf{Q_{y}}$. As
before, the CDW is assumed to appear as soon as the superconducting order
decays due to vortex generation in sufficiently strong magnetic fields. CDW
at the vortex cores should be easily detectable in x\nobreakdash-ray
experiments.

To be specific, the x\nobreakdash-ray scattering intensity is determined by the density-density structure factor, which here reads
\begin{align}
I_{\bq}  = \rho_{\bq} \rho_{-\bq}
\label{Iq}
\end{align}
with~$\rho_{\bq}=\sum_{\mathbf{r}}\exp(-\mathrm{i}\mathbf{q}\mathbf{r})\rho(\mathbf{r})$ the Fourier transform of the charge density.
Herein, the sum is over all lattice sites~$\mathbf{r}$ in the CuO lattice and $\rho(\mathbf{r})$ is equal to $\rho_{\mathrm{CDW}}(\mathbf{r})$,
Eq.~(\ref{rhoCDW}), if $\mathbf{r}$ is a Cu site, and given by
$\rho_{\mathrm{CDW}}(\mathbf{r})+\rho_{\mathrm{QDW},x/y}(\mathbf{r})$, cf. Eq.~(\ref{rhoQDW}),
if $\mathbf{r}$ is an oxygen site on a Cu bond in $x$- or $y$-direction, respectively.

\begin{figure}[t]
\centerline{\includegraphics[width=\linewidth]{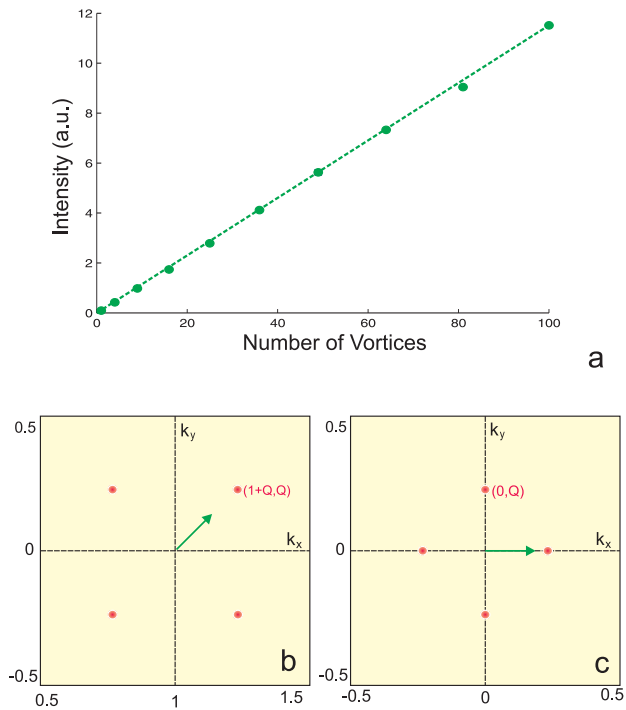}}
\caption{(Color online) (a) Vortex contribution to the intensity of one
 $(200)$ satellite peak. The linear behavior
is clearly visible. (b), (c) Peak structure of the QDW and CDW
order, respectively. The satellites to the (000) and (100) Bragg peaks are
shown.}
\label{Fig3}
\end{figure}

The structure factor~(\ref{Iq}) in the presence of the CDW and QDW charge orders specified by Eqs.~(\ref{rhoQDW}) and~(\ref{rhoCDW})
leads to satellites around the standard Bragg peaks as shown in Fig.~\ref{Fig3}(b) and~(c). Within the $\sigma$-model theory for
the pseudogap state, we expect these charge-order-related peak to appear both at very high magnetic fields~$B>H_{c2}$ so that
the system is in a pure charge order state, and with a lower intensity at intermediate fields $H_{c1}<B<H_{c2}$ where charge order
exists inside the vortex cores. We note that the  QDW only contributes to ``odd'' Bragg peaks, i.e. where the sum of the two integers~$n$
and~$m$ in the reciprocal-lattice vector $\mathbf{K}=(2\pi/a)(n,m)$ is odd. Since QDW order thus grants a signal only at the edge of
the first Brillouin zone, its observation will probably be a challenge.

Finally, let us examine the regime above $H_{c1}$ where many vortices appear
and the vortices form a lattice. The calculation is straightforward; every
vortex has a similar structure to the single vortex. The number of vortices
must be proportional to the applied magnetic field $\mathbf{B}$, and we are
interested in the evolution of the scattering intensity of a charge-order
peak.

The numerical solution is obtained by calculation of the integrated intensity for a
given number of vortices, which are well separated by using Eq.~\ref{Iq}, and
integration of this result over the wave vectors close to the peak.
This procedure results in Fig.~\ref{Fig3}(a), revealing
that the integrated intensity is proportional to the number of vortices,
where the number of vortices is in turn proportional to the applied magnetic
field. Thus, $I_{\mathbf{q}_{0}}^{\mathrm{peak}}\propto |\mathbf{B}|$. This
result is in agreement with a recent experimental work.\cite{chang}

The authors of this study performed a hard-x\nobreakdash-ray experiment and found satellites to the $(2 \ 0 \ 0.5)$
and $(0 \ 2 \ 0.5)$ Bragg peaks that were interpreted as a
result of the formation of CDW order in the CuO planes. At zero field,
the charge order emerges by cooling the system below a critical temperature $%
T_{\mathrm{CDW}} \approx 150 \ $K.

The situation at zero field suggests that this charge signal is a
consequence of the coexistence between superconductivity and charge density wave
order. This is in line with our theoretical finding in Ref.~\onlinecite{mepe2}. On the other hand, the increase of the signal strength by applying the
magnetic field can be attributed to the formation of vortices in the CuO
 plane. Outside the vortices, superconductivity suppresses the charge order
and a field-independent signal is obtained. Inside the vortices, there are
regions that are purely charge-ordered, further enhancing the signal.
Increasing the magnetic field allows more vortices to penetrate the sample
and thus leads to a stronger signal. This can be seen in Fig.~2(b) of Ref.~\onlinecite{chang}. Remarkably, the increase is a linear function of the
applied magnetic field, as suggested.

Further evidence for charge order inside vortex cores was found in Refs.~\onlinecite{julien,hoffman}. The authors of Ref.~\onlinecite{julien} found a
charge-order signal by means of NMR measurement inside the superconducting
phase when a magnetic field is applied. For fixed temperature, the order
starts to give a signal above some threshold field $H_{\mathrm{charge}}$.
Whether the order is uni- or bidirectional was not specified and therefore
it is not completely clear which kind of charge order emerges there, but the
appearance of charge order in the vortex phase confirms our theoretical
finding.

A similar scenario applies to Ref.~\onlinecite{hoffman}, where charge order in the
vortices was found in an STM experiment. It was verified that the order has checkerboard symmetry and therefore fits nicely into our
theoretical findings.

\section{Conclusion}

The $\sigma$-model description\cite{emp} for the pseudogap in the high-$T_c$ cuprates leads to vortices whose geometry may differ from the conventional Ginzburg-Landau picture. The main difference, however, is the onset of charge order in the vortex core, where the superconducting order parameter turns to zero. This gives rise to peaks in the density-density structure factor and explains the CDW signals seen in  x\nobreakdash-ray experiments.\cite{chang, julien, hoffman} We hope that future experimental works will soon clarify the theoretically troubling issue of charge-order  x\nobreakdash-ray peaks along the diagonal of the Brillouin zone.
\newline

\begin{acknowledgments}
K.B.E. and M.E. gratefully acknowledge financial support from the Ministry
of Education and Science of the Russian Federation in the framework of
Increase Competitiveness Program of NUST~\textquotedblleft
MISiS\textquotedblright\ (Nr.~K2-2014-015). Financial support of K.B.E.  and M.E. by SFB/TR12 of DFG is gratefully appreciated. H.M. acknowledges the Yale Prize Postdoctoral Fellowship and C.P. acknowledges the support PALM Labex grant Excelcius.
\end{acknowledgments}

%
%


\begin{thebibliography}{99}
\bibitem{mueller} J. G. Bednorz and K. A. M\"{u}ller, Z. Physik \textbf{64}%
B, 189 (1986).

\bibitem{wise} W. D. Wise, M. C. Boyer, K. Chatterjee, T. Kondo,
T. Takeuchi, H. Ikuta, Y. Wang, and E. W. Hudson, Nature Phys. \textbf{4},
696 (2008).

\bibitem{davis} M. J. Lawler, K. Fujita, J. Lee, A. R. Schmidt, Y. Kohsaka,
C. K. Kim, H. Eisaki, S. Uchida, J. C. Davis, J. P. Sethna, and E.-A. Kim,
Nature (London) \textbf{466 }, 347 (2010).

\bibitem{wu} T. Wu, H. Mayaffre, S. Kr\"{a}mer, M. Horvati\'{c}, C.
Berthier, W. N. Hardy, R. Liang, D. A. Bonn, and M.-H. Julien, Nature (London)
\textbf{477}, 191 (2011).

\bibitem{yazdani} C. V. Parker, P. Aynajian, E. H. da Silva Neto, A. Pushp,
S. Ono, J. Wen, Z. Xu, G. Gu, and A. Yazdani, Nature  (London) \textbf{468}, 677 (2010).


\bibitem{ghiringhelli} G. Ghiringhelli, M. Le Tacon, M. Minola, S.
Blanco-Canosa, C. Mazzoli, N. B. Brookes, G. M. De Luca, A. Frano, D. G.
Hawthorn, F. He, T. Loew, M. Moretti Sala, D. C. Peets, M. Salluzzo, E.
Schierle, R. Sutarto, G. A. Sawatzky, E. Weschke, B. Keimer, and L.
Braicovich, Science \textbf{337}, 821 (2012).

\bibitem{chang} J. Chang, E. Blackburn, A. T. Holmes, N. B. Christensen, J.
Larsen, J. Mesot, R. Liang, D. A. Bonn, W. N. Hardy, A. Watenphul, M.
v. Zimmermann, E. M. Forgan, and S. M. Hayden, Nat. Phys. \textbf{8}, 871
(2012).

\bibitem{achkar} A. J. Achkar, R. Sutarto, X. Mao, F. He, A. Frano, S.
Blanco-Canosa, M. Le Tacon, G. Ghiringhelli, L. Braicovich, M. Minola, M.
Moretti Sala, C. Mazzoli, R. Liang, D. A. Bonn, W. N. Hardy, B. Keimer,
G. A. Sawatzky, and D. G. Hawthorn, Phys. Rev. Lett. \textbf{109}, 167001
(2012).

\bibitem{leboeuf} D. LeBoeuf, S. Kr\"{a}mer, W. N. Hardy, R. Liang, D. A.
Bonn, and C. Proust, Nat.~Phys.~\textbf{9}, 79 (2013).



\bibitem{blackburn} E. Blackburn, J. Chang, M. H\"{u}cker, A. T. Holmes, N.
B. Christensen, R. Liang, D. A. Bonn, W. N. Hardy, U. R\"{u}tt, O.
Gutowski, M. v. Zimmermann, E. M. Forgan, and S. M. Hayden, Phys. Rev. Lett.
\textbf{110}, 137004 (2013).


\bibitem{zaanen} J. Zaanen and O. Gunnarsson, Phys. Rev. B \textbf{40},
7391(R) (1989).


\bibitem{machida} K. Machida, Physica C \textbf{158}, 192 (1989).

\bibitem{poilblanc} D. Poilblanc and T. M. Rice, Phys. Rev. B \textbf{39},
9749 (1989).




\bibitem{tranquada} J. M. Tranquada, B. J. Sternlieb, J. D. Axe, Y.
Nakamura, and S. Uchida, Nature  (London)\textbf{375}, 561 (1995).




\bibitem{emery} V. J. Emery, S. A. Kivelson, and J. M. Tranquada, Proc. Natl.
Acad. Sci. (U.S.A.) \textbf{96}, 8814 (1999).

\bibitem{kivelson} S. A. Kivelson, I. P. Bindloss,  E. Fradkin, V. Oganesyan,
J. M. Tranquada, A. Kapitulnik, and C. Howald, Rev. Mod. Phys. \textbf{75},
1201 (2003).


\bibitem{white} S. R. White and D. J. Scalapino, Phys. Rev. Lett. \textbf{80}%
, 1272 (1998).

\bibitem{hoffman} J. E. Hoffman, E. W. Hudson, K. M. Lang, V. Madhavan, H.
Eisaki, S. Uchida, and J. C. Davis, \emph{Science} \textbf{295}, 466
(2002).



\bibitem{Levy} G. Levy, M. Kugler, A. A. Manuel,  {\O}. Fischer, and M. Li Phys.
Rev. Lett. \textbf{95}, 257005 (2005).

\bibitem{Ishida} K. Matsuba, S. Yoshizawa, Y. Mochizuki , T. Mochiku,
K. Hirata, and N. Nishida, J. Phys. Soc. Jpn. \textbf{76}, 063704 (2007).





\bibitem{cominxray} R. Comin, R. Sutarto, F. He, E. da Silva Neto, L.
Chauviere, A. Frano, R. Liang, W. N. Hardy, D. Bonn, Y. Yoshida, H. Eisaki,
J. E. Hoffman, B. Keimer, G. A. Sawatzky, and A. Damascelli, arXiv:1402.5415.

\bibitem{Taillefer1} N. Doiron-Leyraud, C. Proust, D. LeBoeuf, J. Levallois,
J. Bonnemaison, R. Liang, D. A. Bonn, W. N. Hardy, and L. Taillefer, Nature (London)
\textbf{447}, 565 (2007).

\bibitem{Taillefer2} F. Lalibert\'{e}, J. Chang, N. Doiron-Leyraud, E.
Hassinger, R. Daou, M. Rondeau, B. J. Ramshaw, R. Liang, D. A. Bonn, W. N.
Hardy, S. Pyon, T. Takayama, H. Takagi, I. Sheikin, L. Malone, C. Proust, K.
Behnia, and L. Taillefer, Nat. Commun. \textbf{2}, 432 (2011).

\bibitem{sebastian1} S. E. Sebastian, N. Harrison, R. Liang, D. A. Bonn, W.
N. Hardy, C. H. Mielke, and G. G. Lonzarich, Phys. Rev. Lett. \textbf{108},
196403 (2012).

\bibitem{sebastian2} S. E. Sebastian, N. Harrison, and G. G. Lonzarich,
Rep. Prog. Phys. \textbf{75}, 102501 (2012).






\bibitem{julien} T. Wu, H. Mayaffre, S. Kr\"{a}mer, M. Horvatic, C.
Berthier, P. L. Kuhns, A. P. Reyes, R. Liang, W. N. Hardy, D. A. Bonn, and
M. H. Julien, Nat. Commun. \textbf{4}, 2113 (2013).


\bibitem{emp} K. B. Efetov, H. Meier, and C. P\'{e}pin, Nat. Phys. \textbf{9}, 442 (2013).

\bibitem{mepe} H. Meier, M. Einenkel, C. P\'{e}pin, and K. B. Efetov, Phys.
Rev B \textbf{88}, 020506(R) (2013).

\bibitem{ssscience} L. E. Hayward, D. G. Hawthorn, R. G. Melko, and
S. Sachdev, Science \textbf{343}, 1336 (2014).

\bibitem{sachdevmag} L. E. Hayward, A. J. Achkar, D. G. Hawthorn, R. G.
Melko, and S. Sachdev, arXiv:1406.2694 (2014).



\bibitem{ac} Ar. Abanov and A. V. Chubukov, Phys. Rev. Lett. \textbf{84}, 5608
(2000).

\bibitem{acs} Ar. Abanov, A. V. Chubukov, and J. Schmalian, Adv. Phys. \textbf{52}%
, 119 (2003).

\bibitem{ms2} M. A. Metlitski and S. Sachdev, Phys. Rev. B \textbf{82},
075128 (2010).

\bibitem{mepe2} H. Meier, C. Pepin, M. Einenkel, and K. B. Efetov, Phys.Rev.
B, \textbf{89}, 195115 (2014).

\bibitem{wc} Y. Wang and A. V. Chubukov, arXiv:1401.0712

\bibitem{allais} A. Allais, J. Bauer, and S. Sachdev, arXiv:1402.4807; Ind. J. Phys. 0973-1458 (2014).

\bibitem{chowdhury} D. Chowdhury and S. Sachdev, arXiv:1404.6532.

\bibitem{allais2} A. Allais, D. Chowdhury, and S. Sachdev, arXiv:1406.0503.

\bibitem{blatter} {G. Blatter, M. V. Feigel'man, V. B. Geshkenbein, A. I. Larkin,
and V. M. Vinokur, Rev. Mod. Phys. \textbf{66}, 4 1125--1388} (1994).

\bibitem{tinkham} M. Tinkham, \textit{Introduction to Superconductivity} (McGraw-Hill,
New York, 1975).

\bibitem{cominSTM} R. Comin, A. Frano, M. M. Yee, Y. Yoshida, H. Eisaki, E.
Schierle, E. Weschke, R. Sutarto, F. He, A. Soumyanarayanan, Y. He, M. Le
Tacon, I. S. Elfimov, J. E. Hoffman, G. A. Sawatzky, B. Keimer, and A.
Damascelli, Science, \textbf{343}, 390 (2013).

\bibitem{laplaca} S. Sachdev and R. La Placa, Phys. Rev. Lett. \textbf{111},
027202 (2013).


\end{thebibliography}
\end{document}